\documentclass{aastex63}

\definecolor{boxgray}{HTML}{E0E6E6}
\usepackage{amsmath}
\usepackage{mdframed}
\usepackage{bbm}
\graphicspath{{figures/}}
\newcommand{\fwhm}{\ensuremath{\text{FWHM}_\text{PSF}}}
\renewcommand{\vec}[1]{\ensuremath{\mathbf{#1}}}

\DeclareMathOperator{\sgn}{sign}
\DeclareMathOperator*{\argmax}{arg\,max}

\shorttitle{Star-Galaxy Separation with Gaussian Processes}
\begin{document}

\title{Star-Galaxy Image Separation with Computationally Efficient Gaussian Process Classification}

\correspondingauthor{Amanda L. Muyskens}
\email{muyskens1@llnl.gov}

\author{Amanda L. Muyskens}
\affiliation{Engineering Division, Lawrence Livermore National Laboratory \\
Livermore, CA 94550, USA}

\author{Im\`ene R. Goumiri}
\affiliation{Physics Division, Lawrence Livermore National Laboratory \\
Livermore, CA 94550, USA}

\author{Benjamin W. Priest}
\affiliation{Center for Applied Scientific Computing, Lawrence Livermore National Laboratory \\
Livermore, CA 94550, USA}

\author{Michael D. Schneider}
\affiliation{Physics Division, Lawrence Livermore National Laboratory \\
Livermore, CA 94550, USA}

\author{Robert E. Armstrong}
\affiliation{Physics Division, Lawrence Livermore National Laboratory \\
Livermore, CA 94550, USA}

\author{Jason M. Bernstein}
\affiliation{Engineering Division, Lawrence Livermore National Laboratory \\
Livermore, CA 94550, USA}

\author{Ryan Dana}
\affiliation{Computing Division, Lawrence Livermore National Laboratory\\
Livermore, CA 94550, USA}



% =============================================================================
\begin{abstract}
We introduce a novel method for discerning optical telescope images of stars from those of galaxies using Gaussian processes (GPs).
Although applications of GPs often struggle in high-dimensional data modalities such as optical image classification, we show that a low-dimensional embedding of images into a metric space defined by the principal components of the data suffices to produce high-quality predictions from real large-scale survey data.
We develop a novel method of GP classification hyperparameter training that scales approximately linearly in the number of image observations, which allows for application of GP models to large-size Hyper Suprime-Cam (HSC) Subaru Strategic Program data.
In our experiments we evaluate the performance of a principal component analysis (PCA) embedded GP predictive model against other machine learning algorithms including a convolutional neural network and an image photometric morphology discriminator.
Our analysis shows that our methods compare favorably with current methods in optical image classification while producing posterior distributions from the GP regression that can be used to quantify object classification uncertainty.
We further describe how classification uncertainty can be used to efficiently parse large-scale survey imaging data to produce high-confidence object catalogs.
\end{abstract}


% =============================================================================

\section{Introduction}
\label{sec:intro}

% Need to separate stars from galaxies
% and: lots of data from very faint sources

The production of accurate maps of the cosmos from wide-field optical sky surveys depends fundamentally on our ability to produce reliable catalogs of stars and galaxies from photometric imaging.
Given a pure galaxy catalog, measures of galaxy clustering~\cite{DESJACQUES20181} and gravitational lensing shear~\cite{kilbinger2015cosmology} can inform models of cosmology and galaxy evolution.
However, obtaining competitive model constraints from existing and future surveys requires selecting galaxy samples that span most of the visible sky down to magnitude limits well below that of the best stellar catalogs.
When faint stars are erroneously classified as galaxies, large-angle galaxy clustering statistics can be biased by the low-order angular moments of the Milky Way stellar spatial distribution.
In measuring cosmic shear from galaxies, stellar contamination in the galaxy sample can systematically reduce the large-angle shear correlations and increase the large-angle shear-galaxy cross-correlations, which both bias inferences of the cosmological model.
Cosmic shear also requires accurate knowledge of the point spread function (PSF), typically derived from a sample of stars, that can be biased if misclassified galaxies are used to derive the PSF.
Studies of galactic archeology and discovery of new satellites of the Milky Way are also biased by galaxy sample contamination in the stellar catalog~\cite{Drilica-Wagner2015}.

With the largest-area optical sky surveys such as Pan-STARRS~\cite{flewelling2018pan}, the Dark Energy Survey (DES)~\cite{abbott2019dark}, the Rubin Observatory Legacy Survey of Space and Time (LSST)~\cite{ivezic2019lsst}, or Euclid~\cite{amiaux2012euclid}, star-galaxy mis-classifications can introduce erroneous signals in the search for relativistic clustering~\cite{yoo2014relativistic}, primordial non-Gaussianity~\cite{desjacques2010primordial}, dark energy clustering~\cite{hu2004measuring}, or modified gravity~\cite{renk2016gravity} constraints. Detection of any such effects beyond the standard cosmological model could be among the most impactful scientific results from these surveys if the star-galaxy sample selection can be improved well beyond that shown in recent works~\cite[e.g.,][]{10.1093/pasj/psx126,sevilla2018star}. Excess galaxy clustering detections in the past~\cite[e.g.,][]{sawangwit2011angular,thomas2011excess} are not yet reliable as indicators of new physics without further characterization of systematic errors, including star-galaxy mis-classifications.

Direct methods for star-galaxy image classification based upon image summary information suffice for relatively bright objects.
For instance, the morphological approach~\cite{Vasconcellos11, Slater20} compares the apparent shape of the source with the point spread function (PSF), and is employed in the Dark Energy Survey (DES)~\cite{des2005,abbott2018}.
The Hyper Surpime-Cam (HSC)~\cite{miyazaki2018,Aihara17} survey similarly compares the flux ratio of a galaxy model to a model using the PSF.
Stars and galaxies also exhibit different spectral energy distributions.
\cite{Pollo10} exploit this fact to cluster images in color space based upon the far-infrared data obtained from astronomical surveys.
Many investigators have also used spectral and morphological features to devise supervised learning approaches using decision trees~\cite{vasconcellos2011decision, sevilla2015effect} and ensemble methods such as random forests~\cite{kim2015hybrid}.

% However, modern ground-based astronomical surveys record massive quantities of images, most of which capture faint celestial bodies whose images are almost indistinguishable from point-like sources.\footnote{Although space telescopes can capture higher-resolution images of the same objects, tasking such resources to observe large regions of the sky is prohibitively expensive compared to their gound-based counterparts.}
The simple approaches using hand-crafted features utilized in the past often fail to correctly classify the faintest images in modern surveys, as they comparatively contain less information and more noise than earlier datasets.
Furthermore, the sheer volume of data demands automation with minimal human intervention.
%This combination of factors identifies the star-galaxy separation problem as a rich area of exploitation by representation learning - where we aim to learn the features of the data in addition to preforming prediction.
%Representation learning models include deep neural networks (DNNs) and kernel models such as support vector machines and Gaussian processes (GPs).

% \todo{Mention here how classification labels are generated: space-based surveys such as HST and the future NASA Roman telescop; citizen-science projects like galaxy-zoo.}
Labels for classifying faint star and galaxy images can be obtained from overlapping space-based imaging where the better resolution allows more reliable detection of resolved galaxy features.
However, the Hubble Space Telescope (HST) has a narrow field of view that limits the available labeled data to small subsets of the areas of ground-based surveys. The upcoming ESA Euclid survey will cover a wide area, but will not reach the limiting magnitudes of most ground-based wide-field surveys.
The upcoming NASA Roman survey will reach a limiting magnitude comparable to that of LSST, but will only cover roughly 10\% of the LSST footprint.
Citizen science projects such as Galaxy Zoo can be another source of labeled classifications.

The volume and complexity of the data suggest the use of \emph{representation learning} - where a machine learning model learns an appropriate feature representation in addition to performing prediction.
Deep convolution neural networks (CNNs) are representation learning models that have become very popular in image processing applications due to their ability to capture and exploit local contours in arrays of adjacent pixels \cite{krizhevsky2012imagenet}.
Neural networks (non-convolutional) were previously applied to the star-galaxy classification problem by \citet{odewahn1992automated, sevilla2018star}.
Recently \citet{kim2016star} showed that CNNs compete with random forest classifiers on separating star and galaxy images.
Alternatively, kernel models such as support vector machines (SVMs) and Gaussian processes (GPs) are representation learning models that eschew the explicit construction of a feature map - like CNNs - and instead rely on a kernel, which specifies a similarity function in a representation space that is nonlinearly related to the data space.
\cite{fadely2012star} used SVMs to solve the star-galaxy separation problem, and in \cite{goumiri2020stargalaxy}, GPs are utilized to demonstrate that a low-dimensional embedding of the images result in improved GP classification performance.
While GPs are very attractive due to their fully-Bayesian inference model, the limited expressiveness of popular kernel functions as well as the computational scaling compared to CNNs has slowed their application to high-dimensional domains such as image processing~\cite{bradshaw2017adversarial}.

% our contribution
In this document we demonstrate the viability of GPs as a tool to solve the star-galaxy separation problem by extending the work presented in \citet{goumiri2020stargalaxy}.
We describe a normalization scheme of pre-processing the images.
Then, we overcome the high dimensionality of the data by way of principle component analysis (PCA), dramatically restricting the dimensionality of the structured image data to only its most salient features.
We describe a novel method of GP classification hyperparameter training that exploits sparsity in the procedure of GP prediction that incorporates an approximate nearest neighbor algorithm based on a similar method presented for GP regression in \cite{muyskens2021muygps}.
Using this new method, we show that GPs trained on the embedded data are both computationally efficient and accurate in both small data and large data regimes.
We further show that even when optical data is available in multiple filters, images from a single filter are as good as those produced from the full dataset.
We also show that the joint PCA-GP model compares favorably to other machine learning baselines on both the full and PCA-reduced data.
Further, our method outperforms the morphological discriminator from the HSC pipeline even for very few training images, and demonstrates a large improvement for many training images.
Finally we show how we can make use of the full posterior distribution returned by Gaussian processes to interpret the prediction and quantify uncertainty.
By using this uncertainty, one can identify a small class of less-accurate images that can be verified by non-automated methods.
This process further elevates the performance of the classifier and describes procedure which to parse and label large batches of star-galaxy images reliably incorporating automation.

% =============================================================================

\section{Star-Galaxy Image Data}
\label{sec:data}

The data used in this analysis is from the first public release of the HSC Subaru Strategic Program~\cite{Aihara17}, the deepest ongoing large-scale survey.
In particular, we selected a training set of star and galaxy images from the UltraDeep COSMOS field which is smaller in area but was observed many more times making it significantly deeper than the main survey.
The COSMOS field overlaps space-based higher-resolution imaging from the Hubble Space Telescope (HST) so we cross match the HSC objects to the HST objects that have been labelled as stars or galaxies by~\citet{leauthaud2007cosmos}, who identified stars by looking for the stellar locus in the 2D space of magnitude and peak surface brightness.
While they claim their classification to be reliable up to magnitude $\sim25$, beyond which point sources and galaxies appear identical breaking down their classification, the HSC data is quite a bit deeper ($\sim 27$ in the $i$ filter), so we must be careful in interpreting results at the faint end.

%\begin{figure}
%	\centering
%	\begin{subfigure}[b]{0.49\textwidth}
%		\centering
%		\includegraphics[width=\textwidth]{star}
%		\caption{Star}
%	\end{subfigure}
%	\hfill
%	\begin{subfigure}[b]{0.49\textwidth}
%		\centering
%		\includegraphics[width=\textwidth]{Galaxy}
%		\caption{Galaxy}
%	\end{subfigure}
%	\caption{Example star and galaxy plots with overlaid image (red) and point spread function (blue) for the i band. }
%	\label{fig:explots}
%\end{figure}

	\begin{figure}[!htb]
	\centerline{
		\includegraphics[width=0.9\textwidth]{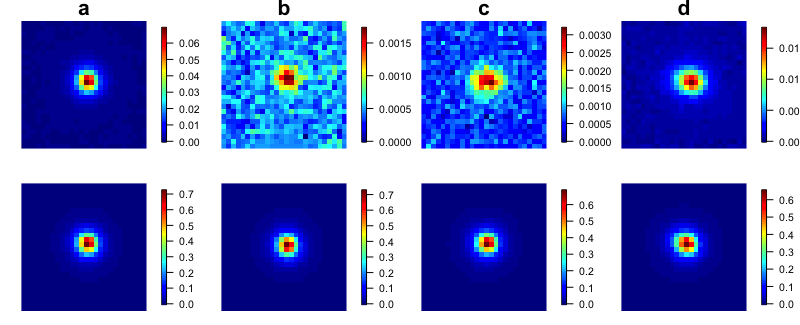}
	}
	\caption{
		Examples of normalized $i$ band images (top row) and their corresponding point spread functions (bottom row).
		a. and b. are star example observations, and c. and d. are galaxy observations. The star examples are more symmetical, but star b. is difficult to distinguish visually from the galaxy example in c. Similarly, galaxy example observation d. although subtly asymmetric, is more similar in appreance to star a. than galaxy example c. 
	}
	\label{fig:stargal}
\end{figure}

Of the five filters, corresponding to different frequency bands that equip HSC, we consider data from the $g$, $r$, $i$, and $z$ band and exclude the $y$ band data because it is significantly shallower than the other bands.
For each filter we extract images from the deblended HSC images, and reject images that have a signal-to-noise ratio lower than 10 in all four bands, so as to remove spurious sources and more closely match the HST catalogs.
Finally, to get rid of remaining artifacts and junk objects, we ensure that the ratio of the $\chi^2$ value for a circular area of interest ($r < 2\,\fwhm$) over the chi squared value of an annulus region around it ($5\,\fwhm < r < 10\,\fwhm$) must be lower than 4, where \fwhm\ is the Full Width at Half Maximum (FWHM) of the point spread function (PSF).
Example images and point spread functions for the $i$ band can be seen in \ref{fig:stargal}.

% =============================================================================

\section{Methodology}
\label{sec:methodology}

% overview of methodology and outline of section
The methodology employed here defines a classifier that assigns images of celestial objects to the ``star'' or ``galaxy'' labels.
The classifier is based upon a workflow that processes images and assigns to them conditional probabilities that they depict stars.
Thresholding this probability obtains the final label prediction.

We describe the full workflow in detail throughout the remainder of this section.
Section~\ref{subsec:pca} describes a normalization and dimensionality reduction procedure using principal component analysis (PCA).
Section~\ref{subsec:gp} defines a GP classifier that accepts this PCA-reduced data and interpolates class labels in terms of a latent predictor.
Section~\ref{subsec:estimation} provides a novel cross-validation-based hyperparameter estimation training procedure for the GP classification model, as well as providing a scalable implementation leveraging the nearest neighbors structure of the data.
Section~\ref{subsec:uq} discusses how to use the GP posterior variance to identify ambiguous images that the model classifies with low confidence.
Finally, we validate our methodology in Section~\ref{sec:results} by training such a classifier on a subset of the pre-labeled images and evaluating its accuracy on the remaining images. 
Further, we compare to baselines and demonstrate the performance in several numerical studies of the data.
% moved CNN + NNGP details to appendix

% splitting the  into disjoint training and test sets.
 % and benchmarked

% good but not sure goes here... merge with above or move?
% We want to train a classifier that turns images of celestial objects into probability distributions of being stars or galaxies.
% Using : the \emph{training set} used for training and the \emph{test set} used for validation and benchmarking.

% move to results section where discussed
% First we use a standard convolutional neural network (CNN) to use as a performance baseline, ignoring the fact that it produces a binary output rather than probability distributions.

\subsection{Normalization and Dimensionality Reduction}
\label{subsec:pca}

% training and test data here...

% Prior to the application of a principal components analysis, it is necessary to normalize the data.
We describe in this section a normalization scheme that produces favorable performance on the star-galaxy classification for all methods, although other choices are possible. 
For a single object, we consider at most 8 images, where the $g$, $r$, $i$, and $z$ bands each contain a photometric image as well as a point spread function (PSF) image produced by the HSC pipeline.
Further, we crop images to the most central $26 \times 26$ pixels per image.
We then independently normalize each of these 8 image classes.
%Each of these 8 classes of images are independently normalized.  

We assume throughout $n$ training examples and $m$ test examples.
Let $W_{j,k}^{(0)} \in \mathbb{R}^{676}$ be the pixel vector defined by the flattened image associated with the $k$th band of the $j$th image, where $k=1, 2, \dots, 8$ and $j=1, 2, \dots, (n + m)$.

%Further, let $W_{k}^0=[W_{1k}^{0T}; W_{2k}^{0T}; \dots; W_{(n+m)k}^{0T}] \in \mathbb{R}^{(n + m) \times 676}$ be the matrix whose rows contain each image associated with channel $k$.
%Then the $j$th flattened vector single image is $W_{jk}^0$ for the $k$th image type.
%Then our normalization is performed in two stages.
We normalize in two stages.
First we remove the background from each image independently,
\begin{equation}
W_{j,k}^1 = W_{j,k}^0 - \min(W_{j,k}^0) \mathbf{1}_{676}.
\end{equation}
Here, $\mathbf{1}_{\ell} = \{1\}^{\ell}$ is the vector of ones.
Then, we normalize the values within each image vector to have at maximum value of 1,
\begin{equation}
W_{j,k} =\frac{W_{j,k}^1} {\max(W_{j,k}^1)}.
\end{equation}
Then we form a full matrix $W \in \mathbb{R}^{(n + m) \times 5408}$ such that
\begin{equation}
W = \begin{bmatrix}
W_{1,1}^T & W_{1,2}^T & ...& W_{1,8}^T\\
W_{2,1}^T & W_{2,2}^T & ...& W_{2,8}^T\\
\vdots & \vdots & \ddots & \vdots\\
W_{(n+m),1}^T & W_{(n+m),2}^T & ...& W_{(n+m),8}^T\\
\end{bmatrix}
\end{equation}

% PCA reduction of data
Next we reduce the dimensionality of the input data by performing a principal components analysis.
For the 8 image classes, we have a total of $8 \times 26^2 = 5,408$ pixels for each object.
We also consider single band images, including both their photometric image and the corresponding PSF image.
These single-band datasets correspond to $(n + m) \times 1352$ vertical slices of $W$.

Both $W$ and its single-band variants are of a significantly higher dimension than is typically considered reasonable to fit with a GP without dimension reduction.
For the rest of this document we will more generally define $W$ to contain any subset of vertical slices of the 8 image types for each object included and therefore have dimensions $(n+m) \times N$.
Then, since $W^T W$ is proportional to the sample covariance matrix of $W$, we perform the eigendecomposition $W^T W= P \Lambda P^T$, where $\Lambda$ is a diagonal matrix that contains the sorted eigenvalues and $P$ is a matrix that contains the corresponding eigenvectors.
Define $P_L$ to be a $N \times L$ matrix comprised of columns of the first $L$ eigenvectors.
Then we can compute our data embedding as $W P_L \in \mathbb{R}^{(n + m) \times L}$.

\begin{figure}[!htb]
	\centerline{
		\includegraphics[width=.95\textwidth]{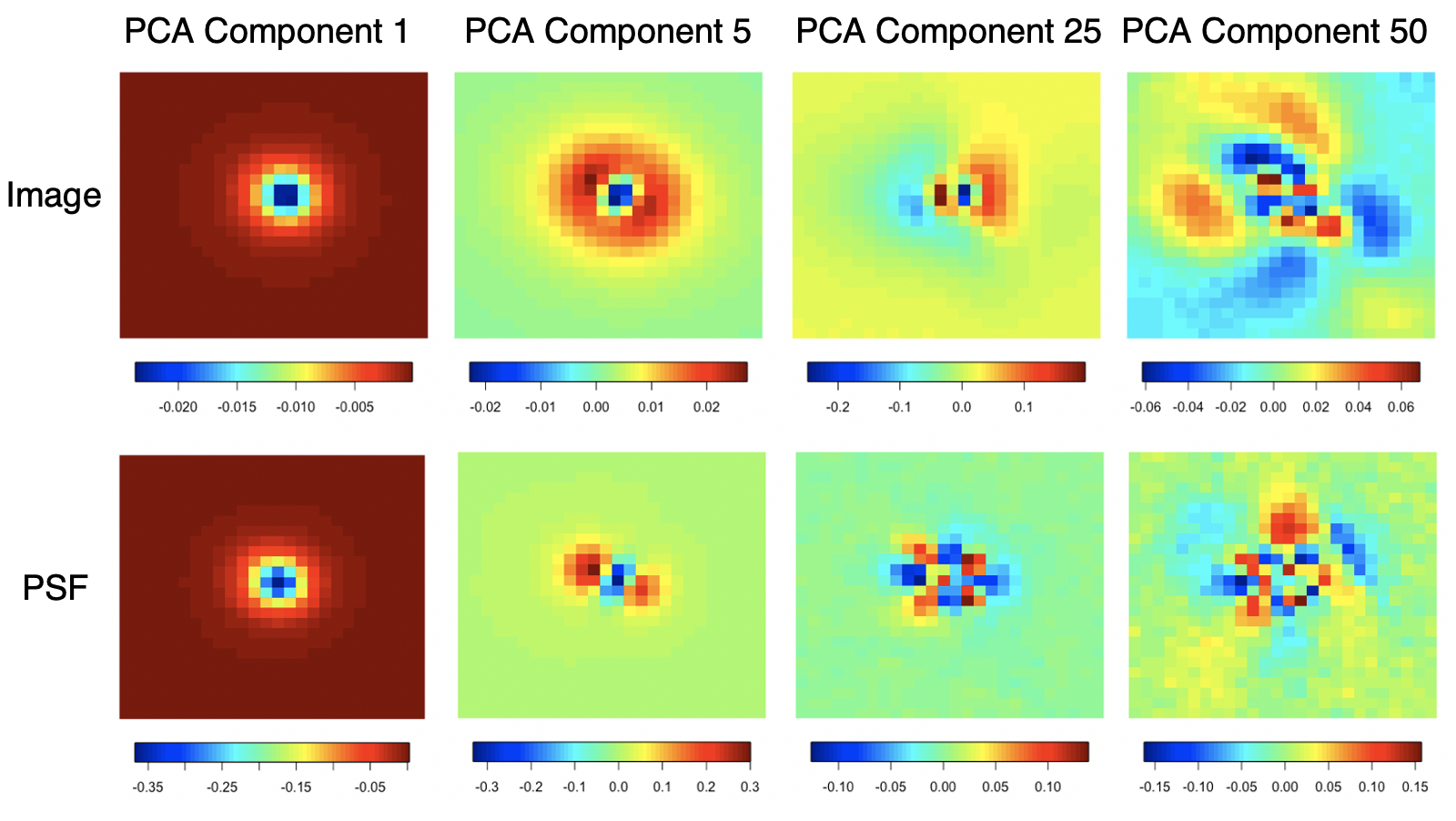}
	}
	\caption{\label{fig:pc}
 PCA component variables for components 1, 5, 25, 50 (left to right) for both the image (top) and point spread function (bottom) for the $i$ channel.
	}
\end{figure}

% how we do PCA in practice and example principal components
Performing the full eigendecomposition of such a large matrix is impractical on a typical computer without some approximation.
We approximate the largest eigenvalues and their corresponding eigenvectors using the methods in \cite{lehoucq1998arpack}.
This is implemented using the ``eigsh'' function in the SciPy Python package~\cite{scipy}.
Figure \ref{fig:pc} shows several example principal components of the image data for the $i$ band images.
We see that the first principal components are symmetrical large-scale features, and as the component number increases, they model non-symmetric, high-frequency features.

% segue to next section

%We found that 30 eigenvalues were sufficient to capture the most important features of the data and obtained a very good prediction accuracy as shown in \autoref{fig:var}.
%Further studies omitted here found similar results as our 30 components model for 50 and 100 eigenvectors.
%At the end of this operation, the reduced-order training set is now a matrix $X$ of size $n \times 30$, and the reduced-order testing set is now a matrix $X_*$ of size $m \times 30$.

% some of this should go in data reduction section?
The matrix $WP_L \in \mathbb{R}^{(n + m) \times L}$ describes $n$ training images and $m$ testing images that have been embedded into $L$ dimensions.
Let $X_\textrm{train} = \{\mathbf{x}_1, \dots, \mathbf{x}_n\}$ be the rows of $WP_L$ that correspond to the $n$ embedded training images.
These images are associated with ground truth class labels $\mathbf{z} \in \{-1, +1\}^n$ such that $\mathbf{z}_i = +1$ when image $\mathbf{x}_i$ is known to be a star and $\mathbf{z}_i = -1$ otherwise.
Similarly, $X^*_\textrm{test} = \{\mathbf{x}^*_1, \dots, \mathbf{x}^*_m\}$ are the rows of $WP_L$ corresponding to the $m$ embedded testing images.
%Define the collected data vector is $\mathbf{y}=(y_1,y_2,...,y_n)^T$ collected at reduced dimension locations matrix $X=(x_1,x_2,...,x_n)^T$.
%The values of $y$ are one-hot encoded to have one value for each classification.
%Typically, these values are chosen to be $-1,1$.

\subsection{Gaussian Processes}
\label{subsec:gp}

% good topic sentence? good overview but not clear how connected to the point of the paper
After the data is embedded on a low-dimensional space, we employ Gaussian Processes (GPs) in order to build a star-galaxy discrimination model.
% gp overview and approach
GPs allow fully-Bayesian inference and quantify uncertainty for a response variable with a Gaussian posterior distribution.
Typically, GPs are used to model continuous response variables, but they can also be used for classification by modeling a latent variable that relates to the different classes, see Chapter 3 of \cite{rasmussen2006gaussian}.
For example, with logistic GP regression the logit of the conditional class probabilities is modeled as a latent Gaussian process.
Here we take a simpler approach where the classes are modeled as the sign of a latent GP.
The challenges are to infer the latent Gaussian process from the training data and to predict the GP, and therefore class label, at a new training point.

% relationship of GP to class labels, and GP technical details
In particular, we employ a GP here as a prior distribution over a discrimination function $f_\theta:\mathbb{R}^L \rightarrow \mathbb{R}$, where $L$ is the embedding dimension of the images described in Section~\ref{subsec:pca}.
Here $\theta$ are hyperparameters that will be defined later.
The GP distribution is described by a mean function, $m(\cdot)$, and a positive definite covariance function, $k_\theta(\cdot, \cdot)$.
The notation $f_\theta \sim \mathcal{GP}(m(\cdot), k_\theta(\cdot, \cdot))$ is used to indicate that a function $f$ has a GP distribution with mean function zero and covariance function $k_\theta$.
By convention, the mean function is assumed to be known and so $m(\cdot) \equiv \mathbf{0}$ is assumed without a loss of generality.

Let $\mathbf{z}=(z_1,\ldots,z_n)'$ denote the known class labels and let $f_\theta(\mathbf{x}_i)$ denote the corresponding value of the latent discriminator at input location $\mathbf{x}_i$.
The class labels are related to the latent GP as
\begin{equation}
\label{eq:thresh}
z_i = \sgn(f_\theta(\mathbf{x}_i))=
	\begin{cases}
		+1 & f_i(\theta) > 0\\
		-1 & f_i(\theta) < 0.
	\end{cases}
\end{equation}
%For the rest of this section we will drop the explicit $\theta$ subscript for clarity, remembering that all quantities are conditioned on the values of $\theta$.
Tuning the hyperparameters $\theta$ conditioned on observations require estimation whose details we defer to Section~\ref{subsec:estimation}.

% We aim to learn a class discrimination function $f:\mathbb{R}^L \rightarrow \mathbb{R}$ on the continuous latent responses so that we can predict the class labels of the test images $X_\textrm{test}$.
% This is accomplished by imposing a GP prior distribution on $f$ and thresholding the posterior mean for $X_\textrm{test}$.
% GP statistical model
We will assume that $\mathbf{f} \in \mathbb{R}^{n}$ constitute evaluations of a continuous, surrogate discrimination function $f_\theta: \mathbb{R}^L \rightarrow \mathbb{R}$ on $X_\textrm{train} = \{\mathbf{x}_1, \dots, \mathbf{x}_n\}$.
Further, we assume that $\mathbf{y}$ are the ``observed'' (prior to class label thresholding by application of Equation~\eqref{eq:thresh}) realizations of $f_\theta$ on $X_\textrm{train}$ perturbed by homoscedastic Gaussian noise $\boldsymbol{\epsilon}$.
% and latent variable $\mathbf{y}$ constitute noisy evaluations of a continuous, surrogate discriminator function $f : \mathbb{R}^L \rightarrow \mathbb{R}$ with homoscedastic noise $\boldsymbol{\epsilon}$, and seek to interpolate $f$'s response $\mathbf{f}_*$ to the unknown testing data $X^*_\textrm{test}$.
We seek to interpolate $f_\theta$'s response $\mathbf{f}_* \in \mathbb{R}^m$ to the unknown testing data $X^*_\textrm{test} = \{\mathbf{x}^*_1, \dots, \mathbf{x}^*_m\}$.
The assumption that $f_\theta \sim \mathcal{GP}(\mathbf{0}, k_\theta(\cdot, \cdot))$ imposes the following Bayesian prior model on $\mathbf{f}$, the true evaluations of $f$ on $X_\textrm{train}$:
\begin{align}
\label{eq:prior_distribution}
\begin{split}
\frac{\mathbf{y}}{\sigma} &= \mathbf{f} + \boldsymbol{\epsilon}, \\
\mathbf{f} &= [f_\theta(\mathbf{x}_1), \dots, f_\theta(\mathbf{x}_n)]^\top \sim \mathcal{N}(\mathbf{0}, K_\mathbf{ff}), \\
\boldsymbol{\epsilon} &\sim \mathcal{N}(0, \tau^2 I_n).
\end{split}
\end{align}
Here $K_\mathbf{ff}$ is an $n \times n$ positive definite covariance matrix on the training data whose $(i, j)$th element is $k_\theta(\mathbf{x}_i, \mathbf{x}_j)$, and $\tau^2$ is the variance of the unbiased homoscedastic noise.
The definition of GP regression then specifies that the joint distribution of all training and testing responses $\mathbf{y}$ and $\mathbf{f}^*$ is given by
\begin{equation}
\label{eq:joint_distribution}
  \begin{bmatrix} \mathbf{y} \\ \mathbf{f}_*
  \end{bmatrix}
  = \mathcal{N} \left ( 0, \sigma^2
  \begin{bmatrix}
    K_\mathbf{ff} + \tau^2 I_n & K_\mathbf{f*} \\
    K_\mathbf{*f} & K_{**}
  \end{bmatrix}
      \right ).
\end{equation}
Here $K_\mathbf{f*} = K^\top_\mathbf{*f}$ is the cross-covariance matrix between the training and testing data;
that is, the $(i,j)$th element of $K_{f*}$ is $k(\mathbf{x}_i, \mathbf{x}^*_j)$.
Similarly, $K_\mathbf{**}$ is the covariance matrix of the testing data, and has $(i, j)$th element $k_\theta(\mathbf{x}^*_i, \mathbf{x}^*_j)$.
Finally, we are able to compute the posterior distribution of the testing response $\mathbf{f}^*$ on $X^*_\textrm{test}$ as
\begin{align}
\label{eq:posterior_distribution}
\begin{split}
  \mathbf{f}^* \mid X_\textrm{train}, X^*_\textrm{test}, \mathbf{y} &\sim \mathcal{N}(\bar{\mathbf{f}}^*, \sigma^2 C), \\
  \bar{\mathbf{f}}^* &\equiv K_\mathbf{*f} (K_\mathbf{ff} + \tau^2 I_n)^{-1} \mathbf{y}, \\
  C &\equiv K_{**} - K_\mathbf{*f} (K_\mathbf{ff} + \tau^2 I_n)^{-1} K_\mathbf{f*}.
\end{split}
\end{align}

\begin{figure}[!htb]
	\centerline{
		\includegraphics[width=0.7\textwidth]{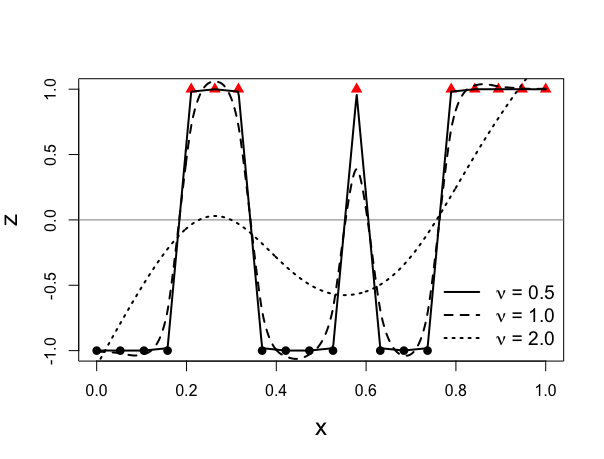}
	}
	\caption{
	1-D example of our GP classifier with various Mat\'ern kernel hyperparameters.
	Lines represent $\bar{\mathbf{f}}^* $ and $X$ observations where $\bar{\mathbf{f}}^* <0$ are classified as one class (galaxy), and those with $\bar{\mathbf{f}}^* >0$ are the other (star).
	}
	\label{fig:ex_gp}
\end{figure}

Equation~\eqref{eq:posterior_distribution} gives the posterior mean $\bar{\mathbf{f}}^*$ in closed form.
However, this equation depends upon $\mathbf{y}$, the hypothetically observed perturbed realizations of the surrogate model $f$, whereas we only have access to the thresholded class indicators $\mathbf{z}$.
Some sources such as \cite{rasmussen2006gaussian} estimate $\mathbf{y}$ using a Laplace approximation.
However, this approximation requires repeated covariance matrix solves that are impractical with large training data.
Accordingly, we identify $\mathbf{z}$ for $\mathbf{y}$ in Equation~\eqref{eq:posterior_distribution} and use the $\{-1, +1\}$ labels directly as regression targets, which is similar to least-squares GP classification (see Chapter 6.5 of \cite{rasmussen2006gaussian}).
Identifying $\mathbf{y}$ with $\mathbf{z}$ gives us an alternative solution for the posterior mean $\bar{\mathbf{f}}^*$:
\begin{equation} \label{eq:posterior_mean}
\bar{\mathbf{f}}^* \equiv K_\mathbf{*f} (K_\mathbf{ff} + \tau^2 I_n)^{-1} \mathbf{z}.
\end{equation}
We obtain $\bar{\mathbf{f}}^*$ using Equation~\eqref{eq:posterior_mean} and reverse the latent variable encoding using Equation~\eqref{eq:thresh}, learning the predicted class labels of the training data $\mathbf{z}^* \in \{-1, +1\}^m$.
%Then for computational reasons, we substitute observed $z$ for $\mathbf{y}$ in the above equations.
Demonstrations of the classification induced by this substitution in a one-dimensional numerical example is in Figure \ref{fig:ex_gp}.
%For testing image $i$, we set $\mathbf{z}^*_i = +1$ if $\bar{\mathbf{f}^*}_i > 0$, and set $\mathbf{z}^*_i = -1$ otherwise.
The $i$th test image corresponding to input $\mathbf{x}^*_i$ is classified as a star if the associated latent GP mean is positive and is classified as a galaxy otherwise.
Section \ref{subsec:uq} will additionally describe using the posterior variance to identify images that the model predicts with low confidence, leading to uncertain class predictions.

% Matern kernel
The posterior distribution given in Equations~\eqref{eq:posterior_distribution} and \eqref{eq:posterior_mean} depends on the choice of kernel function.
We select the Mat\'ern kernel, which is a stationary and isotropic kernel that is commonly used in the spatial statistics GP literature due to its flexibility and favorable properties \cite{stein2012interpolation}.
In particular, the Mat\'ern kernel is used here since it allows greater control over the smoothness of the GP than the commonly-used radial basis function kernel.
A general expression for the kernel is
\begin{equation}
k_\text{Mat\'ern}(\vec{x}, \vec{x^\prime}) =  \sigma^2 \frac{2^{1-\nu}}{\Gamma(\nu)}
	{\left(
		\sqrt{2\nu} \frac{{\| \vec{x} - \vec{x^\prime} \|}_2^2}{\ell}
	\right)}^\nu
	K_\nu{\left(
		\sqrt{2\nu} \frac{{\| \vec{x} - \vec{x^\prime} \|}_2^2}{\ell}
	\right)},
\label{eq:matern}
\end{equation}
where $\nu>0$ is a smoothness parameter, $\ell>0$ is a correlation-length scale hyperparameter, $\sigma^2>0$ is a scale parameter, $\Gamma$ is the Gamma function, and $K_\nu(\cdot)$ is a modified Bessel function of the second kind.
We will optimize these hyperparameters in our analysis rather than fixing them for convenience.
Furthermore, using the previously discussed normalization and dimensionality reduction procedures, the authors observed similar performance from this kernel, the radial basis function kernel, and the neural network Gaussian process (NNGP) kernel \cite{yang2019fine}, but these results are omitted.

% segue to next section
Given a set of hyperparameters, predictions can be obtained as described in Section \ref{subsec:gp}.
However, the hyperparameters should be selected from the data to provide optimal predictions.
The next section discusses a cross-validation-based approach to hyperparameter estimation that is computationally efficient for large training data sizes.

\subsection{Hyperparameter Estimation}
\label{subsec:estimation}

% intro to hyperparameter estimation and difficulties
GP hyperparameters are typically estimated using a likelihood-based approach such as maximum likelihood estimation or Bayesian calibration.
However, the task is challenging in part because evaluating the likelihood is computationally expensive ($O(n^3)$) and the optimization problem is non-convex.
Cross-validation using a grid search can be employed, but prediction still involves forming the kernel matrix and inverting a large matrix, which can also be prohibitively expensive, and limit the optimized parameter values.
Models such as \cite{fuentes2001high} and \cite{gramacy2007tgp} seek to improve computational efficiency in hyperparameter optimization by partitioning the domain and asserting independence over the partitions.
However, the partitioning can lead to discontinuous predictions and the hyperparameter estimates can be poor if there is significant correlation across partition boundaries.
Other methods make low-rank (\cite{banerjee2008gaussian}) or sparsifying assumptions (\cite{kaufman2008covariance}) that allow for faster approximate estimation.
A comprehensive review of computationally inexpensive GP-like methods are in \cite{heaton2019case}.

% local kriging to get around difficulties
Models such as \cite{gramacy2015local}, utilize only local training observations to fit independent GP models for each prediction location using maximum likelihood estimation.
This method is extremely computationally efficient and parallelizable, but the estimates can be inaccurate with only a few correlated local observations to estimate a set of hyperparameters.
We improve on this type of local estimation method in two primary ways.
First, we avoid maximum likelihood in favor of cross validation because the kriging weights ($K_\mathbf{*f} (K_\mathbf{ff} + \tau^2 I_n)^{-1} $) are sparser than the correlation function \eqref{eq:matern}.
Therefore, the predictions using only local data is an improved approximation over a local likelihood approximation.
Second, because we assume an overall stationary GP model, we borrow information across spatial locations to improve robustness of the hyperparameter estimates.
Further, in a stationary GP, a new model is not necessary for each prediction location so a large number of prediction can be made efficiently.
This method is a novel extension of the approximate leave-one-out cross-validation approach for GP regression in \cite{muyskens2021muygps} extended to the more complex GP classification estimation problem.

% leave-one-out classification accuracy
Cross-validation seeks parameter values to maximize out-of-sample classification accuracy.
To formally describe the procedure, let $\theta$ denote the hyperparameters that require estimation and $k_\theta(\cdot,\cdot)$ a GP kernel of interest.
In the  Mat\'ern kernel, define $\theta=(\sigma^2,  \nu, \ell, \tau^2)^T$.
Here we omit estimation of $\sigma^2$ because predictions of $f_i(\theta)$ do not depend on  $\sigma^2$.
Therefore, a different estimation method to define the uncertainty quantification of the classification prediction will be used outlined in the next section.
Let $\hat{z}_i$ be the $i$th leave-one-out class label prediction of $z_i$ given the set of all training points excluding $\mathbf{x}_i$.
$\hat{z}_i$, is obtained in two steps.
First, let $\mathbf{x}_{-i}=\{\mathbf{x}_1, \mathbf{x}_2, \dots, \mathbf{x}_{i-1}, \mathbf{x}_{i+1}, \dots, \mathbf{x}_n\}$ and $\mathbf{z}_{-i}= \{z_1,z_2,\dots,z_{i-1},z_{i+1},\dots,z_n\}$ be the observations and labels excluding the $i$th training point.
Then we modify Equation~\ref{eq:posterior_mean}  to obtain the mean GP prediction by regressing the training labels on the corresponding inputs,
\begin{equation}
\label{cvpred1}
%\hat{f}(\mathbf{x}_i) = \bar{f}_i = K_\theta(\mathbf{x}_i, \mathbf{x}_{-i}) K_\theta(\mathbf{x}_{-i},\mathbf{x}_{-i})^{-1} \mathbf{z}_{-i},
\hat{f}_\theta(\mathbf{x}_i) = \bar{f}_i = K_{i,-i} K_{-i.-i}^{-1} \mathbf{z}_{-i}.
\end{equation}
Here $K_{i, -i}$ is the cross-covariance between $\mathbf{x}_i$ and $\mathbf{x}_{-i}$, while $K_{-i, -i}$ is the covariance among points in $\mathbf{x}_{-i}$, both in terms of $k_\theta(\cdot, \cdot)$ as in Equation~\eqref{eq:joint_distribution}.
%where $\mathbf{x}_{-i}=(\mathbf{x}_1, \mathbf{x}_2, \dots, \mathbf{x}_{i-1}, \mathbf{x}_{i+1}, \dots, \mathbf{x}_n)^T$ and $\mathbf{z}_{-i}$ is the vector of observations excluding the $i$th observation, or $\mathbf{z}_{-i}=(z_1,z_2,\dots,z_{i-1},z_{i+1},\dots,z_n)^T$.
We then use Equation~\ref{eq:thresh} to round $\hat{f}_\theta(\mathbf{x}_i)$ to the nearest class label $\hat{z}_i$.
Note that the GP classifier can be interpreted as a special case of a technique called kernel classification, but the direct correspondence is not given here to avoid unnecessary equations.
%This value is then rounded to the nearest possible class label value, so that
%%
%\begin{equation}
%\label{cvpred2}
%z_i(\theta) = \sgn\left(f_i(\theta)\right)=\begin{cases}
%		+1 & f_i(\theta) > 0\\
%		-1 & f_i(\theta) < 0.
%	\end{cases}
%\end{equation}
%
%Note that the sign function transformation is needed so that the GP prediction, $f_i(\theta)$, is in $\{-1,1\}$.

Although many criterion for classification accuracy are possible, we select the cross-entropy loss, also referred to as the log loss. 
Define $\delta : \mathbb{R} \rightarrow [0,1]^2$ as a ``padded'' softmax function, where $\delta(a) = \left(\frac{e^a}{e^a + e^{-a}}, \frac{e^{-a}}{e^a + e^{-a}} \right )$.
$\delta$ takes a point prediction in $\mathbb{R}$ and converts it into a probability distribution.
Then the cross entropy loss is
\begin{equation}
\label{loss}
%Q(\theta) =-\sum_{i=1}^{n} \biggl\{z_i log\left[f_i(\theta)^{\star}\right]+ (1-z_i)log\left[1-f_i(\theta)^{\star}\right]\biggr\}.
Q(\theta) =-\sum_{i=1}^{n} \biggl\{\frac{z_i + 1}{2} \log\left[ \delta \left( f_\theta (\mathbf{x}_i) \right )_0 \right]+ \left ( 1-\frac{z_i + 1}{2} \right ) \log\left[ \delta \left( f_\theta (\mathbf{x}_i) \right )_1 \right]\biggr\}.
\end{equation}
The hyperparameters are estimated as the maximizer of the leave-one-out cross-validation accuracy
\begin{equation}
\label{thetahat}
\hat{\theta}=\argmax_{\theta} -Q(\theta).
\end{equation}
The loss function \eqref{loss} is maximized using the "optimize" function from the SciPy Python package, but other choices are possible.

% % intuition behind procedure
% Our GP classifier can be interpreted as a special case of a technique called kernel classification.
% The kernel classifier prediction for the $i$th class label is
% %
% \begin{equation}
% \label{kernclass}
% \hat{z}_i=\sgn\left(K_{i,-i} \Omega \mathbf{z}_{-i}\right),
% \end{equation}
% %
% where $\Omega$ is a diagonal matrix with $i$th diagonal element denoted $\omega_i$.
% The GP class-label prediction in \eqref{cvpred1} is then a special case of \eqref{kernclass} if $\omega_i$ is the $i$th coefficient of $\mathbf{z}_{-i}$ in \eqref{cvpred1}.
% In particular, the $i$th diagonal element of the weight matrix $\Omega$ is the sum of the $i$th column of the matrix $K_{-i,-i}^{-1}$ in \eqref{cvpred1}.

% nearest neighbors cross-validation
When there are a large number of observations, the cross-validation optimization is more computationally expensive than traditional maximum likelihood estimation;
the procedures have complexity $O(n^4)$ and $O(n^3)$, respectively.
Hence, local kriging is used in place of full kriging, so that only the $q$ nearest neighbors to the prediction location are considered, where $q \ll n$.
The complexity of the local kriging approach is $O(n q^3)$, which is linear in the sample size $n$ and tractable compared to the cubic scaling of likelihood-based optimization approaches.

Let $\mathbf{x}_{N_i}$ be the set of training observations nearest to $\mathbf{x}_i$, and let $\mathbf{z}_{N_i}$ be their corresponding labels.
Similar to \eqref{cvpred1}, the $i$th nearest neighbors GP prediction is
\begin{equation}
f^{NN}_\theta(\mathbf{x}_i) = K_{i,N_i} K_{N_i, N_i}^{-1} \mathbf{z}_{N_i},
\end{equation}
where $K_{i, N_i}$ is the cross-covariance between $\mathbf{x}_i$ and $\mathbf{x}_{N_i}$, while $K_{N_i, N_i}$ is the covariance among points in $\mathbf{x}_{N_i}$, similar to Equations~\eqref{eq:joint_distribution} and \eqref{cvpred1}.
%
%where $\mathbf{x}_{N_i}$ is the set of observations nearest to $\mathbf{x}_i$ and $\mathbf{z}_{N_i}$ is the corresponding vector of responses.
Hyperparameter optimization proceeds following Equation~\eqref{thetahat}, substituting $f^{NN}_\theta(\mathbf{x}_i)$ for $f_\theta(\mathbf{x}_i)$.
%We then use Equation~\eqref{eq:thresh} to round $f^{NN}_\theta$ to a class label prediction, $\hat{z}_i^{NN}$, and the hyperparameter optimization proceeds following \eqref{thetahat}.

% batching to improve computational efficiency
Batching, a common technique utilized in machine learning, allows us to obtain further computational efficiency.
In traditional batching, random locations are selected individually or in groups from the training data.
The loss function in \eqref{loss} is then replaced by a summation over a subset of the training data whose indices are collected in a set denoted $B$.
That is, the loss function obtained with batching is
\begin{equation}
\label{lossB}
Q_B(\theta) =-\sum_{i \in B} \biggl\{\frac{z_i + 1}{2} \log\left[ \delta \left( f_\theta^{NN} (\mathbf{x}_i) \right )_0 \right]+ \left ( 1 - \frac{z_i + 1}{2} \right ) \log\left[ \delta \left( f_\theta^{NN} (\mathbf{x}_i) \right )_1 \right]\biggr\}.
%Q_B(\theta)=-\sum_{i\in B}  \biggl\{z_i log\left[f_i(\theta)^{\star}\right]+ (1-z_i)log\left[1-f_i(\theta)^{\star}\right]\biggr\},
\end{equation}
where the hyper-parameter estimate is obtained by maximizing this function as in \eqref{thetahat}.
This approximation introduces some variability into the optimization problem because the batch indices are randomly selected, but the computational savings can be significant if $|B| \ll n$.
If $n$ is small, then there is little value to be gained with batching and all of the data can be used for hyperparameter estimation.

% batching details
We conclude our discussion on hyperparameter estimation with a comment on how batching is performed here within the nearest neighbors framework.
If batched observations are randomly selected from the data and only this data is used in training, the nearest neighbors of each batched observation will be artificially distant from one another.
This structure does not well-represent the densely-sampled dataset, and therefore the hyperparameter selection would be biased in fitting low-frequency correlations, but be unable to model the more important high-frequency correlations.
To avoid this problem, all points in the original training set are considered as possible nearest neighbors to the points included in the batch.
Hence, more data is ultimately used to estimate the hyperparameters than is included directly in the batch.
Furthermore, points are excluded from being included in a batch if all their neighbors are the same class since that point will have the same contribution to the loss function regardless of the hyperparameters or kernel choice.

% non-stationary comment
% If a block of points is selected and the data is actually non-stationary across the domain, the batching will swing widely and possibly have poor convergence.

% segue to 'UQ'
Given a trained GP classifier, class label predictions can be made for a new image.
However, in addition to a class label prediction, the GP also outputs a Gaussian distribution that contains information about the confidence of the prediction.
We continue with a discussion of how to use this information to assess whether an image is a star or galaxy with confidence, or whether the image class is ambiguous.

\subsection{Ambiguous Classification} % new name?
\label{subsec:uq}

Although uncertainty quantification for predictions is defined in Gaussian processes through conditional distributions, it is unclear how to utilize this latent information for classification uncertainty quantification.
We describe a method that utilizes the latent uncertainties through the context of statistical hypothesis testing. Using this method, we include a third classification label we refer to as "ambiguous" where the latent Gaussian process is non-significantly different than the class cutoff value, which is 0 in our case.
By removing these images from the star and galaxy designations, we are able to improve performance of the of the classifier, and identify the images that may need manual review, further imaging, or other verification methods or increase our reliability in the object's classification.

For demonstration, assume that if the $i$th training observation is a galaxy, $\mathbf{z}_i= -1$ and if it is a star $\mathbf{z}_i= +1$. Then if on the test set, the latent predictions are less than 0, we predict it is a galaxy, and if it is greater than 0, a star.

Consider the hypotheses
\begin{equation}
\begin{split}
H_0: f_i = 0 \\
H_1: f_i \neq 0
\end{split}
\end{equation}
Ideally, if $f_i<0$, this would correctly identify the $i$th observation as a galaxy, and we would hope to reject $H_0$ and accept the alternative.
This would mean that we are confident in the correct identification of the observation as a galaxy.
Inversely, if $f_i>0$, we would hope to fail to reject $H_0$ since this is the incorrect conclusion.
In this testing formulation, a Type I error ($\alpha$) is when 0 is not contained in the interval, but the classification is incorrect.
 Similarly, a Type II error ($1-\beta$) is when 0 is contained in the interval, but the prediction would be in the correct classification.
We can formally test this hypothesis by creating latent prediction intervals given the Gaussian process standard error.
For each testing location, we compute the following prediction interval using local observations
\begin{equation}
f_\theta(\mathbf{x}_i) \pm 1.96 * \widehat{\sigma}^2 C_{ii},
\end{equation}
where $C_{ii}$ is defined as the element in the $i^{th}$ row and $i^{th}$ column of $C$ defined in Equation~\eqref{eq:posterior_distribution} with optimized hyperparameter values.
If this interval contains 0, we classify it as ambiguous, but otherwise use the classification given by the sign of the prediction as previously described.

However, the computation of these intervals requires an estimate of the prediction variance $\sigma^2$ we were unable to estimate via cross-validation.
After the other hyperparameters in $\theta$ have been estimated as in the previous section, we compute the above intervals for leave-one-out predictions using local predictions.
Then, we minimize a fixed function of the Type I ($\alpha$) and Type II ($1-\beta$) errors in these tests to estimate $\sigma^2$.

Generally as the number of Type I errors increase, the number of Type II errors decrease and vice versa.
Note that because there are many more correct classifications in our method, estimation of the Type II error is more accurate than that of the Type I error, but a Type I error is considered more severe.
Defining a category of "ambiguous" images is ultimately a balance of these errors,
Further, the more ambiguously classified images, the higher the accuracy of the remaining classification.
However, this requires manual review of more ambiguous images.
Therefore, based on available resources, different choices may be made as to how many images could reasonably be classified as such.
We present several example functions that yield a range of possible values.
\begin{enumerate}
	\item $\alpha+1-\beta$.
	\item $2\alpha+1-\beta$.
	\item $4\alpha+1-\beta$.
	\item $10\alpha+1-\beta$.
	\item $n_{i}* \alpha+(1-\beta)*n_{c}$,
\end{enumerate}
where $n_c$ is the number of correctly identified images, and $n_i$ is the number of incorrectly classified images.
Other choices are possible, but we found minimizing these functions gave a variety of estimates of $\sigma^2$.

%From a practical standpoint, it may be difficult to define an exact function of errors a researcher is willing to accept.
%Instead, it may be the case that there are human resources to manually check a fixed number of automated object classifications.
%We use the uncertainty quantification from our GP method to order the observations by the certainty of the prediction, and therefore identify the objects that should be manually reviewed for more accurate classification.
%Define the absolute normalized prediction for the $i$th object as
%\begin{equation}
%\frac{|\hat{z}_i|}{\widehat{\sigma} \sqrt{C_{ii}}}.
%\end{equation}
%These statistics are then ordered from smallest to largest.
%This orders the predictions of the objects from least certain to most certain.
%Then, the first fixed number of observations are those to which manual review will yield the biggest improvement in classification.
%These methods of uncertainty quantification is demonstrated on our data in the next section.
%
%% =============================================================================

\section{Results and Discussion}
\label{sec:results}

In this section, we perform a serious of numerical studies to evaluate the accuracy of our method in the classification of stars and galaxies.
Throughout this section, we randomly sample training and testing images from our available dataset so that there are equal numbers of star and galaxy images. In total, the dataset has 31,798 labeled objects, where 13,133 objects are stars, and 18,665 objects are galaxies.
In all studies, we allow the testing dataset to be 1,000 images, equally sampled from the available stars and galaxies. The primary metric for success we report is the accuracy of the metric:
\begin{equation}
\frac{1}{1000}\sum_{i=1}^{1000} \mathbbm{1} \{\hat{z_i}=z_i\},
\end{equation}
where $\mathbbm{1}$ is an indicator function that is 1 when the classification of the model matches the true classification and 0 otherwise.
This out-of-sample testing data accuracy procedure is then repeated for 100 simulation iterations where the training and testing samples are randomly re-drawn, normalized, embedded, and the classifiers are retrained independently in each iteration.

\begin{figure}[!htb]
	\centerline{
		\includegraphics[width=.7\textwidth]{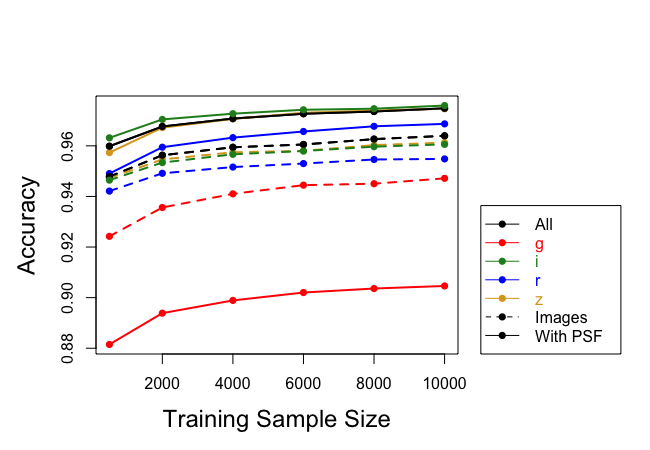}
	}
	\caption{\label{fig:channels}
		The \textit{i}-band images produce the best prediction. The \textit{g}-band channel produces the worst classifications. In all other cases, including the PSF images in addition to the primary star or galaxy image improves classification accuracy.
	}
\end{figure}

First, we formally study the influence utilizing data from various bands and point spread images in the classification of stars and galaxies. 
For training set sizes $\{500,2000,4000,6000,8000,10000\}$, we compute the accuracy of a classifier where the PCA reduction is performed on various subsets of the total available images of each object.
The results of this simulation are in Figure \ref{fig:channels}. These results demonstrate that $g$ and only classification is significantly worse than all other data configurations.
In all other bands, including the point spread function images improves the accuracy of the classifiers. 
Overall, using only the $i$ band produces the most accurate predictions. 
This is not surpising as HSC prioritized the $i$ band when data collection was good. 
Although the magnitude of this improvement is small, we conclude that data from this band is sufficient. 
Therefore, in all other numerical studies, we consider data only from this band.

\begin{figure}[!htb]
	\centerline{
		\includegraphics[width=0.9\textwidth]{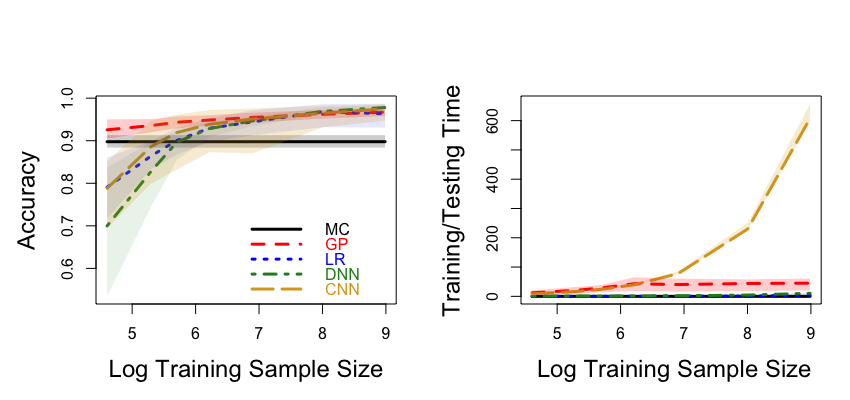}
	}
	\caption{\label{fig:cnn_comparison}
		Left: Star-galaxy classification accuracy versus training sample size. Several classification algorithms are shown by the different colors and line types. See the text for the description of each algorithm.
		Right: Computational runtimes for the combined operations of training and testing the classifiers as measured on a personal computer on a single core. 
		The Mat\'{e}rn kernel is more accurate than CNNs for small sample sizes. Both methods linearly scale, but the slope of the CNN computation is lower than that of the Mat\'{e}rn.
	}
\end{figure}

Next, we compare our method to other machine learning methods as well as the {\it extendedness} morphological classifier from the HSC pipeline~\cite{Bosch18}. We test the accuracy of these classifiers for a wide range of training sample sizes.
In comparison to our GP method, we first consider the morphological classifier. 
It classifies stars and galaxies based on the magnitude difference between a PSF and galaxy model. 
For a star the difference should be zero, while a galaxy will have a significant deviation from zero. 
A hard cut of $mag_{PSF}-mag_{galaxy} = 0.0164$ is used to set the boundary between the two cases. 
While the choice of cut was not optimized for any particular science use case, it gives a baseline to compare against.

Next we consider a convolution neural network classification model. 
This is the generally accepted state-of-the-art machine learning model for image classification tasks. 
We implement this method via the ``Keras" Python package with the architecture give in the Appendix.
We utilize a similar cross-entropy loss function as our Gaussian process method, with 30 epochs. 
To provide other comparisons, we consider both a logistic regression (LR) and dense neural network (DNN) applied to the same  PCA reduced data fit by our GP method.

\begin{figure}[!htb]
	\centerline{
		\includegraphics[width=.5\textwidth]{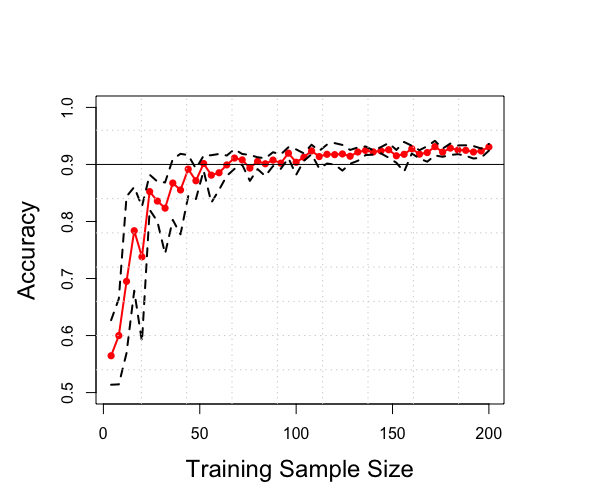}
	}
	\caption{\label{fig:small}
	Extremely small sample GP classifier accuracy results and 90\% empirical confidence intervals over 100 simulation iterations.
	The smallest sample size is $n=4$, where only 2 star and 2 galaxy observations are provided.
	The mean accuracy of the GP classifer exceeds the morphological classifier accuracy of 0.9 at around 80 training observations.
	}
\end{figure}

The results of this study are in Figure \ref{fig:cnn_comparison} plotted on the log scale of sample size.
The morphological classifier is constant in training size because it does not rely on training data.
In low training sample sizes, all methods other than the GP classifier perform worse than the morphological classifier.
However, in training sample sizes, all machine learning methods perform significantly better than the morphological classifier (approximately 0.08 improvement), and the differences between the methods is small.
In terms of training computing time, the CNN is by far much more expensive than the other methods.
Although the training time of our GP classifier is more than that of the other faster methods, the scaling is much improved over the traditional $O(n^3)$ scaling.
In order to exlore the minimum sample size needed to outperform the morphological classifier, we explore the accuracy of our method with extremely small sample sizes in Figure \ref{fig:small}.
The mean accuracy of the GP classifer exceeds the morphological classifier accuracy of 0.9 at around 80 training observations.
This means that only approximately 0.3\% of the total 31,798 object observations need to be labeled in order to demonstrate classification improvement on the morhpological classifier with our GP method.

\begin{figure}[!htb]
	\centerline{
		\includegraphics[width=0.8\textwidth]{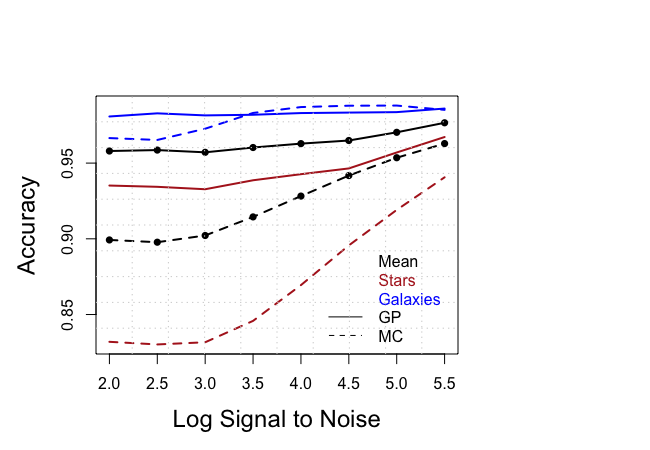}
	}
	\caption{\label{fig:sn_comparison}
	Comparison of Gaussian process (GP) and the morphological classifier (MC) over signal to noise ratios broken out by star and galaxy accuracy. "Mean" represents the accuracy reported in other analyses, which is the mean performance across stars and galaxies.
	}
\end{figure}

We further compare our GP classifier and the morphological classifier in Figure \ref{fig:sn_comparison}.
For each noted signal-to-noise ratio, we excluded images that were below that thresholded value and estimated the accuracy using only 1,000 training star-galaxy images randomly sampled from the images of sufficient signal.
Although the morphological classifier does well at identifying galaxies, in low signal-to-noise images, it does a poor job identifying stars.
Our GP method performs well at identifying for both star and galaxy images even in low signal-to-noise images.

Finally, we summarize results of our uncertainty quantification in Table \ref{tab:uq}. 
In all cases, we demonstrate improved star-galaxy classified accuracy by determining between approximately 4 and 21 images as ambiguous out of a total of 1,000 testing images. 
Although the accuracy of the ambiguously classified testing images was better than 0.5, its average performance is significantly below that of the star-galaxy classified images.
In future piplines that automate the star-galaxy designation for large amounts of data, employing this method would allow for improvement in accurate classification by indicating the "ambiguous" images that should be verified by a human operator. 
This represents a significant improvement over using the same human operator resources to verify the classification of random images in the prediction set. 

\begin{figure}[!htb]
	\centerline{
		\includegraphics[width=.5\textwidth]{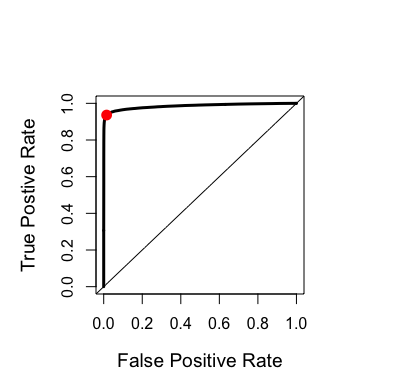}
	}
	\caption{\label{fig:roc}
		ROC curve of our GP classifier with 5,000 training observations averaged over 100 simulation iterations. Red point represents our classifier with 0 threshold, and area under the curve is 0.985.
	}
\end{figure}

As a continuation of our uncertainty analysis, we demonstrate traditional classification of the receiver operating characteristic (ROC) curve in Figure \ref{fig:roc}.
In these plots, the true positive rate and false positive rates are compared for a variety of cutoff values to compare to the $\bar{\mathbf{f}}^* $ predictions.
In this case, define the true positive rate to be the proportion of galaxies that are correctly classified as such, and the false positive rate as the 1- the proportion of stars that are correctly classified.
A perfect classifier would have a true positive rate of 1 and a false positive rate of 0, which would yield an "area under the curve” or AUC of 1.
Averaged across 100 simulation iterations for 5,000 training samples, the AUC is estimated to be approximately 0.985 in Figure \ref{fig:roc}.

% latex table generated in R 3.6.0 by xtable 1.8-4 package
% Tue Feb  9 14:43:21 2021
\begin{table}[ht]
	\centering
	\begin{tabular}{rrrrrrr}
		\hline
		$\widehat{\sigma^2}$ Optimizer	&	& n=1,000 & n=5,000 & n=10,000& n=15,000 & n=20,000\\ 
		\hline
		&	\# Ambiguous & 4.3 & 4.2 & 5.3 & 4.5 & 5.5 \\ 
		$\alpha+1-\beta$ &	Ambiguous Accuracy & 0.641 & 0.681 & 0.792 & 0.742 & 0.790 \\ 
		&	Classified Accuracy & 0.954 & 0.965 & 0.967 & 0.968 & 0.970 \\ 
		\hline
		&	\# Ambiguous& 19.1 & 10.8 & 13.0 & 10.9 & 15.0 \\ 
		$2\alpha+1-\beta$&	Ambiguous Accuracy & 0.748 & 0.753 & 0.790 & 0.763 & 0.788 \\ 
		&	Classified Accuracy & 0.956 & 0.966 & 0.968 & 0.969 & 0.972 \\ 
		\hline
		&	\# Ambiguous & 21.0 & 11.0 & 13.0& 11.1 & 15.1 \\ 
		$4\alpha+1-\beta$&	Ambiguous Accuracy & 0.754 & 0.754 & 0.791 & 0.761 & 0.789 \\ 
		&	Classified Accuracy & 0.956 & 0.966 & 0.968 & 0.969 & 0.972 \\ 
		\hline
		&\# Ambiguous & 21.0 & 11.0 & 13.0& 11.1 & 15.1 \\ 
		$10\alpha+1-\beta$ &	Ambiguous Accuracy & 0.753 & 0.754 & 0.791 & 0.761 & 0.789 \\ 
		&	Classified Accuracy & 0.956 & 0.966 & 0.968 & 0.969 & 0.972 \\ 
		\hline
		&	\# Ambiguous & 5.8 & 6.8 & 6.1 & 5.5 & 6.7 \\ 
		$n_{i}* \alpha+(1-\beta)*n_{c}$&	Ambiguous Accuracy & 0.628 & 0.711 & 0.780 & 0.739 & 0.798 \\ 
		&	Classified Accuracy & 0.954 & 0.966 & 0.967 & 0.968 & 0.970 \\ 
		\hline
		&		All testing Observations Accuracy & 0.952 & 0.964 & 0.966 & 0.967 & 0.969 \\ 
		\hline
	\end{tabular}
	\label{tab:uq}
	\caption{Results from ambiguous classification via Gaussian Process uncertainties for 1,000 testing observations.}
\end{table}

% =============================================================================

\section{Conclusion}
\label{sec:conclusion}

We have shown that applying Gaussian processes, in particular with a Mat\'ern kernel, to reduced-order images of stars and galaxies allows classifying images with great accuracy and efficiency.
Our classification method outperforms standard methods like morphological classifiers as well as other machine learning methods such as CNNs, while providing a higher degree of interpretability of the predictions and requiring fewer training samples.
We have first demonstrated that by normalizing and PCA embedding the data, we can apply GP weights to interpolate masked classification predictions on images of new objects.
These results could inform future astronomical image processing strategies and software pipelines for large sky surveys in the future.

We developed a novel method of GP classification hyperparameter estimation that allows for nearly linear order of estimation (as opposed to traditional $O(n^3)$).
This allows our methods to be efficiently trained on more images than traditional maximum likelihood methods, allowing our methods to use the labeled resources of the large Rubin Observatory data to apply to new large-scale studies.
Since our methods parallelize computation into small, local computations, there is opportunity in future work to scale up these methods on HPC systems.

In our studies, we consider balanced training and testing data, where there are equal numbers of stars and galaxies in our samples.
In this case, our assumption of a mean-zero GP is reasonable, since we have equal $z=[-1, 1]$ samples.
However, to train on all labeled star-galaxy images for future surveys, these assumptions must be relaxed.
Further research must be done in order to explore behavior of this classifier in the unbalanced training data case, and whether a change in mean function could compensate for this effect.
Further, our models consider only the isotropic, stationary Mat\'ern kernel.
Future work could consider anisotropic effects where distances in the PCA components are treated differently.
Future work could also be pursued in generalizing our scalable estimation method to non-stationary through a non-stationary mean or kernel function.
Finally, although we demonstrate the effectiveness and efficiency of our classification method on star-galaxy separation, this method could be more generally applied to other classification problems in astronomy and other fields.
For example, one could consider applying these methods to other space situational awareness applications such as non-resolved object characterization.

% =============================================================================

\acknowledgments

This work was performed under the auspices of the U.S. Department of Energy by Lawrence Livermore National Laboratory under Contract DE-AC52-07NA27344 with IM release number LLNL-JRNL-821985.
Funding for this work was provided by LLNL Laboratory Directed Research and Development grant 19-SI-004.

This document was prepared as an account of work sponsored by an agency of the United States government. Neither the United States government nor Lawrence Livermore National Security, LLC, nor any of their employees makes any warranty, expressed or implied, or assumes any legal liability or responsibility for the accuracy, completeness, or usefulness of any information, apparatus, product, or process disclosed, or represents that its use would not infringe privately owned rights. Reference herein to any specific commercial product, process, or service by trade name, trademark, manufacturer, or otherwise does not necessarily constitute or imply its endorsement, recommendation, or favoring by the United States government or Lawrence Livermore National Security, LLC. The views and opinions of authors expressed herein do not necessarily state or reflect those of the United States government or Lawrence Livermore National Security, LLC, and shall not be used for advertising or product endorsement  purposes.


%\vspace{5mm}
%\facilities{HST(STIS), Swift(XRT and UVOT), AAVSO, CTIO:1.3m,


%\software{astropy \citep{2013A&A...558A..33A},  
%          Cloudy \citep{2013RMxAA..49..137F},
%          SExtractor \citep{1996A&AS..117..393B}
%          }

%\newpage % added this

\bibliographystyle{aasjournal}
\bibliography{Star-Galaxy-Kernels}

\appendix
	\subsection{Convolutional Neural Network}
\label{app:cnn}

Convolutional neural networks are a class of deep neural networks which are commonly used for image classification, where neurons in hidden layers are only connected to a localized subset of neurons from the previous layer.

% CNN architecture and training
The CNN analysis was implemented in R using the Keras package.
The network architecture summary provided by the package is shown below.
The two dropout layers have rates of 20\%.
The convolutional and dense layers utilize the ReLU activation functions, and the output layer employs the softmax activation function so that the output is a probability value in $(0,1)$.
Table \ref{tab:cnntraining} provides details on the CNN training.

\begin{table}[ht]
	\centering
	\begin{tabular}{|l|l|}\hline
		Setting       & Value    \\ \hline\hline
		Learning Rate & 1  \\
		Validation Split & 0.2   \\
		Decay Factor  & 0.95 \\
		% Learning Rate Schedule & ?  \\ not relevant for adadelta?
		Optimizer     & Adam\\
		Loss Function & Categorical cross-entropy \\
		% Momentum      & ? \\ not relevant for adadelta?
		Batch Size    & 256  \\
		Epochs        & 30\\\hline
	\end{tabular}
	\caption{Details on CNN training.}
	\label{tab:cnntraining}
\end{table}

\begin{verbatim}
	Model: "sequential_3"
	___________________________________________________________________________________________
	Layer (type)                            Output Shape                         Param #       
	===========================================================================================
	conv3d_5 (Conv3D)                       (None, 26, 26, 2, 32)                544           
	___________________________________________________________________________________________
	max_pooling3d_5 (MaxPooling3D)          (None, 6, 6, 2, 32)                  0             
	___________________________________________________________________________________________
	conv3d_4 (Conv3D)                       (None, 6, 6, 2, 64)                  18496         
	___________________________________________________________________________________________
	max_pooling3d_4 (MaxPooling3D)          (None, 2, 2, 2, 64)                  0             
	___________________________________________________________________________________________
	flatten_13 (Flatten)                    (None, 512)                          0             
	___________________________________________________________________________________________
	dense_59 (Dense)                        (None, 400)                          205200        
	___________________________________________________________________________________________
	dropout_29 (Dropout)                    (None, 400)                          0             
	___________________________________________________________________________________________
	dense_58 (Dense)                        (None, 200)                          80200         
	___________________________________________________________________________________________
	dropout_28 (Dropout)                    (None, 200)                          0             
	___________________________________________________________________________________________
	dense_57 (Dense)                        (None, 100)                          20100         
	___________________________________________________________________________________________
	dense_56 (Dense)                        (None, 2)                            202           
	===========================================================================================
	Total params: 324,742
	Trainable params: 324,742
	Non-trainable params: 0
	___________________________________________________________________________________________
\end{verbatim}
% =============================================================================

\end{document}